# Intrinsic Perturbation of the Landau Levels in Metals and Semiconductors at Low Temperatures

A. M. Awobode

Department of Physics, University of Illinois, Urbana-Champaign, IL 61801, USA

**Abstract**

It is shown that the frequency of the de Haas–van Alphen effect in non-superconducting metals at very low temperatures is significantly corrected by a perturbative term which appears in the Landau equation sequel to an extension of the Pauli equation. The correction to the frequency (or period) of the de Haas–van Alphen oscillations is found to depend on the Fermi energy and the measurable anomalous part of the electron gyro-magnetic factor. Furthermore, it is shown that as a consequence of the perturbing term the electronic specific heat $C_v$ of a dilute, degenerate Fermi gas, under high magnetic fields (B > 25 Tesla) and at ultra-low temperatures (T ~ $10^{-3}$K) shows an anomalous behavior, and at a finite temperature becomes vanishingly small, i.e $C_v \approx 0$ as the temperature approaches absolute zero. Precision measurement at low temperatures and high magnetic fields of the magneto-optical absorption in simple-band semiconductors is suggested as an immediate way of detecting the modification of the Landau levels due to the weak perturbation term which corrects in a magnetic field, the kinetic energy of the electrons.

PACS number(s): 71.70.Di, 71.10.Ca, 78.20.-e

## 1. Introduction: The Landau Equation and Quantum Oscillations

Quantum magnetic oscillations in metals [1], the integral/fractional Quantum Hall effects in two-dimensional electron gas [2,3,4] and the giant Faraday rotation in magnetic semiconductors [5] are a few of the fascinating phenomena which occur at low temperatures and/or high magnetic fields. They are due, in essence, to the existence of quantized Landau levels and are a manifestation of the behavior of systems of electrons under extreme thermodynamic conditions and reduced dimensionality. We shall here consider in the following, new observations and phenomena which may occur when the temperature is further lowered, the magnetic field intensity is significantly increased or the concentration of electrons is reduced. The results here presented are consequences of an intrinsic contribution to the Hamiltonian of the non-relativistic electron in a magnetic field. An extra kinetic energy term (analogous to the anomalous contribution to the electron spin magnetic moment) corrects the Landau equation, which below is briefly reviewed in relation with the Dirac electron theory.

In the non-relativistic limit, the Dirac equation when coupled to the electromagnetic field (**A**, $\varphi$) reduces (in the lowest-order approximation) to the well-known Pauli equation written as follows;

$$E_n \psi_n = \left[ \frac{\hat{\boldsymbol{\pi}}^2}{2m} + e\varphi - \frac{e\hbar g_s}{2mc}(\hat{\mathbf{S}} \cdot \mathbf{B}) \right] \psi_n \; ; \; \hat{\boldsymbol{\pi}} = \hat{\mathbf{p}} - e\mathbf{A}/c, \; \hat{\mathbf{S}} = \frac{\hbar}{2}\hat{\boldsymbol{\sigma}} \qquad (1.1)$$

where $\hat{\mathbf{S}}$ is the spin, e is the electronic charge, $\hat{\mathbf{p}}$ is the canonical momentum operator[1], $\hat{\boldsymbol{\pi}}$ is the kinetic momentum $m\hat{\mathbf{v}}$ and $g_s = 2$ is the spin gyromagnetic factor. Putting $\hat{\mathbf{S}} = 0$ and $\varphi = 0$, the above equation is transformed to that of Landau, i.e.,

$$E_n \psi_n = \frac{\hat{\boldsymbol{\pi}}^2}{2m} \psi_n \qquad (1.2)$$

In three-dimensions, the eigenvalues of equation (1.2) are given as

$$E_n(k_z) = \frac{\hbar eB}{mc}(n+1/2) + \frac{\hbar^2 k_z^2}{2m}; \qquad (n = 0,1,2,...) \qquad (1.3)$$

where $k_z$ is the wave vector in the z-direction, c is the speed of light and $\hbar$ is related to Planck's constant $h$; the eigenfunctions $\psi_n$ are expressed, in terms of the Hermite polynomials $\mathcal{H}_n$, as

$$\psi_n(x,y,z) = \exp\{i(kx+kz)\} \mathcal{H}_n[\alpha(y-y_0)] \exp\{-\alpha^2(y-y_0)^2/2\} \qquad (1.4)$$

where $(x,y,z)$ are the particle coordinates, $\alpha = (|e|B/\hbar c)^{1/2}$ is the reciprocal of the characteristic magnetic length $a_m$ and $y_0 = -\hbar k_x c/eB = -k_x a_m^2$ is the center of the Hermite polynomial $\mathcal{H}_n$. The Landau levels $E_n$ are highly degenerate due to n and $k_z$, either of which may be fixed, thus reducing considerably the degeneracy. Also, the energy levels described by (1.3) are equally spaced parabolas with respect to $k_z$; the wave-vector $k_z$ is quantized because the system, in three dimensions, is confined in the z-direction.

An extension of the Pauli equation arising from a modified Dirac equation (which describes the evolution of the particle wavefunction) coupled to the electromagnetic field has been described [6, 7]. In the corrected Pauli equation the coefficients of the operators $\hat{\mathbf{S}} \cdot \mathbf{B}$ and $\hat{\mathbf{L}} \cdot \mathbf{B}$ (where $\hat{\mathbf{L}}$ is the orbital angular momentum vector) yields the corrections $\delta_S$ and $\delta_L$ to the gyromagnetic factors $g_S$ and $g_L$ respectively [8]. The corrected spin g is in reasonable agreement with experiments, while both the $g_S$ and $g_L$ calculated from the corrected Pauli equation are compatible with the experimental data obtained by Kusch et al who measured the quantity $\delta_S - 2\delta_L$ in atomic systems [9]. The correction to $g_S$ is accurately measured and widely known; new precision measurements will confirm in detail, the correction to the factor $g_L$.

We observe that, as a consequence of the extension of the Pauli equation, the Landau equation (1.2) will also be modified by the appearance of an extra term. It will be shown in the following section that the new term, considered as a perturbation, corrects in a simple way, the energy levels of electrons placed in a static magnetic field. As a result, interesting electronic properties such as the frequency of oscillation (with the reciprocal of the magnetic field $1/B$) of the magnetization, the electrical and the thermal conductivities etc., will be modified. Also, the thermodynamic properties of a dilute, electron gas in a magnetic field are significantly altered as a consequence. We note Kohn's proof [10] that, down to the lowest temperatures, the cyclotron frequency of an electron in a magnetic field is unchanged by electron-electron interactions etc.

---

[1] Notation: Vector-operators - bold font with cap ($\hat{\mathbf{w}}$), ordinary vectors - bold font (**w**) and scalars - text (w).

## 2a. Perturbation of the Landau Levels

We consider the extended Pauli equation obtained as a non-relativistic limit of the corrected Dirac equation (see **Appendix II** for a complete derivation):

$$E_n \psi_n = \left[ \frac{\hat{\pi}^2}{2m} - \frac{\hat{\pi}^2 \hbar^2 m^2 \varepsilon^2 \bar{\alpha}^2}{2m(4)^2 p_n^4} - \frac{eg}{2mc}\left(1 + \frac{\hbar^2 m^2 \varepsilon^2 \bar{\alpha}^2}{4^2 p_n^4}\right)(\hat{\mathbf{S}} \cdot \mathbf{B}) \right] \psi_n \; ; \; \hat{\mathbf{S}} = \frac{\hbar}{2}\hat{\sigma} \quad (2.1)$$

The quantity $\delta(p_n) \equiv \hbar^2 m^2 \varepsilon^2 \bar{\alpha}^2 / 4^2 p_n^4 = \lambda(mh\varepsilon\bar{\alpha}/p_n^2)^2$, where $\lambda \approx 1.6 \times 10^{-3}$ and $p_n$ is the eigenvalue (corresponding to the state $\psi_n$) of the free-particle momentum operator $\hat{\mathbf{p}}$. Also, $\bar{\alpha} \equiv [1 - e\mathbf{A} \cdot \mathbf{p}/p^2 c + (eA/pc)^2]^{-1}$. For $\hat{\mathbf{S}} = 0$ (or $\hat{\mathbf{S}} \cdot \mathbf{B} = 0$), eqn (2.1) reduces to

$$E_n \psi_n = \left[ \frac{\hat{\pi}^2}{2m} - \frac{\hat{\pi}^2 \delta(p_n)}{2m} \right] \psi_n \quad (2.2)$$

which may also be written as

$$(H_0 - H_1)\psi_n = E_n \psi_n \quad (2.3)$$

where $H_0 = \hat{\pi}^2 / 2m$ is the unperturbed Landau Hamiltonian eqn(1.2) and $H_1 = H_0 \delta(p)$ is the perturbation term in which $\delta(p_n) = \lambda(h\varepsilon/mv_n^2)^2$, where we have expressed in terms of the free-particle velocity $v_n$ the momentum eigenvalue $p_n = mv_n$ and put $\bar{\alpha} \approx 1$ for $p \gg eA/c$; the perturbation series may therefore be expanded in the dimensionless parameter $\lambda$. Thus the corrected energy level $E'_n$, as a consequence of the intrinsic perturbation $H_1$, is given by the series:

$$E'_n = E^{(0)}_n - E^{(1)}_n \lambda + E^{(2)}_n \lambda^2 - E^{(3)}_n \lambda^3 + \cdots, \quad (2.4)$$

where $E^{(j)}$ is the order of the energy correction and $E^{(0)}_n = <n|H_0|n> \equiv E_n$. Also $E^{(1)}_n = <n|H_0 a^*|n>$, hence $E^{(1)}_n = E^{(0)}_n a^* \equiv E_n a^*$ where the quantity $a^* \equiv (h\varepsilon/mv_n^2)^2$.

The Landau levels are highly degenerate, and by the well-known Brillouin-Wigner time-independent degenerate perturbation theory, the j-th order correction to the energy is, in general, given by the expression;

$$E^{(j)}_n = <n|H_0(R_n H_0)^{j-1}|n> \; ; \; R_n = \sum_{m \neq n} \frac{|m><m|}{E_n - E_m} \quad (2.5)$$

The Hamiltonian $H_0$ is Hermitian; therefore its eigenfunctions are orthonormal. Hence, all terms in eqn (2.5) with $j \geq 2$ vanish altogether, thereby reducing (2.4) to the simple form $E'_n - E_n = -E_n \delta(p_n)$, from which follows the expression,

$$-\Delta E_n / E_n = \delta(p_n) \; ; \; \delta \neq 0 \quad (2.6)$$

where $\Delta E = (E'_n - E_n)$. Thus, there is a downward energy shift $\Delta E_n$, which is directly proportional to the energy of a given Landau level $E_n$. As a consequence the spacing between the Landau levels is decreased by a factor $[1 - \delta(p)]$ which manifests as an increase in the density of states.

Also, the corrected eigenfunction $|\psi_n> = |n> + R_n H_1 |n> + (R_n H_1)^2 |n> + ...$ is such that only the first term is non-zero, which implies that $|\psi_n> = |n>$ for all n; that is, the states are uncorrected by the perturbation. This is consistent with the fact that $H_0$ and $H = H_0 - H_0 \delta(p)$ commute, and thus have common eigenfunctions which are similarly degenerate. Let $|n>$ be an arbitrary common eigenvector of $H$ & $H_0$; it follows that $H|n>$ gives, as in eqn (2.6), $E'_n = E_n - E_n \delta(p_n)$. Also, the perturbed Hamiltonian $H$ and $H_0$ have the same symmetry and therefore the states are all shifted in energy leaving the degeneracy intact.

Similar results are again obtained if, instead of the above arguments, we put (for convenience) $m' \equiv m[1-\delta(p)]^{-1}$ where $m$ is the measured rest mass determined by experiments based on established phenomena such as the Compton scattering of weakly-bound electrons. Substituting the "effective mass" $m'$ into $E'_n \psi_n = (\hat{\pi}^2 / 2m')\psi_n$ gives, for each Landau level, the same expression $E'_n = E_n - E_n \delta(p_n)$ as written above. Specifically, in zero external field (**A** = 0) the "effective mass" $m'_F$ is $m'_F = m[1 - (h\varepsilon\lambda / mv^2_F)]^{-1}$ and, neglecting Coulomb interactions, the corrected Fermi energy for a system of electrons is

$$E'_F = \hbar^2 k_F^2 / 2m' = E_F - E_F \delta(p_F) \qquad (2.7)$$

Hence, the Fermi surface contracts as a consequence of the interaction from which has originated the correction to the electron g-factors. We note also that eqn(2.1) for any state, in the absence of external fields, becomes $E\vartheta(\mathbf{r}) = [\hat{\mathbf{p}}^2(1-\delta(p))/2m]\vartheta(\mathbf{r})$ which reduces, in the limit of low velocities, to the free-particle Schrodinger equation $E\vartheta(\mathbf{r}) = (\hat{\mathbf{p}}^2 / 2m)\vartheta(\mathbf{r})$ because, as previously shown [6], $h\varepsilon \to 0$ as $p \to 0$.

**Many-electron systems in a Magnetic Field:** The perturbed Landau equation described above refers to the energy of a single particle. In order to relate the preceding results to the electron gas, it is necessary to discuss their application to many-particle systems. Thus we write,

$$\hat{H}_i = \frac{\pi^2_i}{2m}\left[1 - \left(\frac{h\varepsilon}{mv^2_{i,\sigma}}\right)^2 \lambda\right] \quad \text{and} \quad \hat{H} = \sum_i H_i \quad ; \quad \lambda \approx 1.6 \times 10^{-3} \qquad (2.8)$$

where $\hat{H}_i$ is the Hamiltonian of the $i^{th}$ electron and $\hat{H}$ is that of the system of particles with eigenvalue $E_n^{(N)}$ corresponding to the N-body eigenstate $\psi_n^{(N)}$. It follows therefore that the grand canonical partition function Z is

$$Z(V,T,\mu) \equiv Tr e^{-\beta(\hat{H}-\mu\hat{N})} = \sum_{N,n} e^{-\beta(E_n^{(N)}-\mu N)} \qquad (2.9)$$

where $\hat{N}$ is the number operator and $\mu$ is the chemical potential of the system. Hence, all thermo-dynamical properties which can be derived from the partition function Z will reflect the correction to the Landau levels.

## 3. Corrected Landau Levels in Metals and Semiconductors

The typical atomic energy level is of the order of 1eV to 10eV; hence the effect of the intrinsic perturbation term as described by eqn (2.6) may escape detection in atoms, except by means of exceedingly high precision measurements of the energy levels or the Zeeman shift in noble gases at very low temperatures and high magnetic fields. However, it may be possible to observe the influence of the perturbing term $H_1$ by considering the magnetic energy of electrons in solids (metals and semiconductor). Examples of typical phenomena in which the intrinsic perturbation of the Landau levels can produce significant effects are the de Haas–van Alphen oscillations in metals, the specific heat of a dilute degenerate Fermi gas in a magnetic field and the magneto-optical properties of semiconductors. We shall here consider first the de Haas-van Alphen effect.

**A. The de Haas–van Alphen Effect:** The oscillation of the magnetization **M** (or the magnetic susceptibility) as a function of the field, i.e., the de Haas-van Alphen effect can be influenced by the correction described in eqn (2.7). The well-known, quantum mechanical method of describing the effect, first demonstrated by Landau for the free electron gas [11], later developed by Onsager [12] and finally generalized by Lifshitz & Kosevich, gives for the magnetization [13],

$$M_{osc} = -A(B,\eta)\sin\left[\frac{c\hbar S_m(\eta)}{2\pi eB} \pm \frac{\pi}{4} - \pi\right] \; ; \; B \ll \frac{\pi ckT}{e\hbar}\frac{\partial S}{\partial E} \qquad (3.1)$$

where $S_m$ is any extremal cross-sectional area of the Fermi surface in a plane normal to the magnetic field, and the amplitude of the oscillations from which the $dS_m/dE$ may be determined is,

$$A(B,\eta) = \frac{4V}{\hbar^3(2\pi)^{3/2}}\left(\frac{e\hbar}{c}\right)^{1/2}\frac{\beta S_m(\eta)e^{-\lambda}}{|\partial^2 S(\eta,p_z)/\partial p_z^2|^{1/2}_m B^{1/2}}\cos\left[\frac{1}{2m_o}\frac{dS_m(\eta)}{d\eta}\right]$$

Now, if $E \to E' = E + q$ (where $q \ll E$ is a small correction) and $p_z = const.$, then to first order,

$$S_m(E';p_z) = S_m(E;p_z) + q\frac{dS_m}{dE} \qquad (3.2)$$

If $q = -E_F\delta$ as described in eqn(2.7), where $E_F \equiv \eta$ is the Fermi energy in the absence of the magnetic field, then

$$S_m(\eta';p_z) = S_m(\eta;p_z) - \eta\delta\frac{dS_m}{dE}\Big|_\eta$$

Substituting the above into the expression for $M_{osc}$, we find that the argument of the sine function (and therefore the frequency of the de Haas–van Alphen oscillations) has changed as given by the expression

$$M_{osc}' = -A(B,p_z)\sin\left[\frac{c\hbar S_m(\eta)}{2\pi eB} - \frac{\eta c\delta\hbar}{2\pi eB}\frac{dS_m}{dE}\Big|_\eta \pm \frac{\pi}{4} + \pi\right] \qquad (3.3)$$

which shows that the frequency of the oscillation as function of $1/B$ has been corrected by the second term in the square brackets as follows;

$$\Delta f = \frac{-c\eta\delta\hbar}{2\pi e}\frac{dS_m}{dE} \quad (3.4)$$

The frequency correction $\Delta f$ depends on both the Fermi energy $\eta$ and $(dS_m/dE)|_\eta$ which may vanish (or change magnitude) as some parameter of the system is modified and/or an external field is varied. Alternatively, we can express the change in terms of the period,

$$T' = \Delta(1/B) = 2\pi\left[\frac{c\hbar S_m(\eta)}{2\pi e} - \frac{\eta c\delta\hbar}{2\pi e}\frac{dS_m}{dE}\right]^{-1} \quad (3.5)$$

For a gas of free electrons, the Fermi surface is spherical and $dS_m/dE = 2\pi m^*/\hbar^2$ where $m^*$ is the cyclotron effective mass of the electron; the correction is about four or five orders of magnitude smaller than the observed frequency i.e $f \sim 10^{13}$, $\Delta f \sim 10^8$. If however, band structure effects are taken into consideration and $(dS_m/dE)$ is large at an extremal point on the Fermi surface, then the corrections to $f$ and $\Delta(1/B)$ described above may become large enough to be clearly observed at low temperatures of the order $10^{-3}$ K.

**B. The Electronic Specific Heat at Low Temperatures:** For a three-dimensional system of electrons in a volume $V_s$ placed in magnetic field, the total energy is

$$E_T = \frac{V_s m^* \omega_c}{2\pi^2 \hbar}\sum_{n=0}^{\infty}\int_{-\infty}^{\infty}\frac{E_n(k_z)\,dk_z}{\exp[\beta(E_n-\mu)]+1} \quad ; \quad \beta = \frac{1}{k_B T} \quad (3.6)$$

Also, the total number of electrons $N$, from which the chemical potential $\mu(T,B)$ may be determined, is expressed as

$$N = \frac{V_s m^* \omega_c}{2\pi^2 \hbar}\sum_{n=0}^{\infty}\int_{-\infty}^{\infty}\frac{dk_z}{\exp[\beta(E-\mu)]+1} \quad (3.7)$$

where $D(B) \equiv V_s m^* \omega_c / 2\pi^2 \hbar$ in (3.6) and (3.7) is the number of states per unit length of the Landau cylinder.

We can get an indication of the consequences of the intrinsic perturbation if we substitute into (3.6) & (3.7) $E \to E'$ where from (2.7) $E'_n = E_n - E_n\delta$. We observe that from the equation

$$E_T = \frac{V_s m^* \omega_c}{2\pi^2 \hbar}\sum_{n=0}^{\infty}\int_{-\infty}^{\infty}\frac{(E_n(k_z)-E_n\delta)\,dk_z}{e^{-\beta E_n\delta}\,e^{\beta(E_n-\mu^*(T,B))}+1} \quad ; \quad \delta \gtrsim 1\times 10^{-3} \quad (3.8)$$

and the other from which the chemical potential $\mu^*(T,B;\delta)$ may be determined:

$$N = \frac{V_s m^* \omega'_c}{2\pi^2 \hbar}\sum_{n=0}^{\infty}\int_{-\infty}^{\infty}\frac{dk_z}{e^{\beta(aE_n-\mu^*(T,B))}+1} \quad ; \quad a = (1-\delta), \quad (3.9)$$

that the exponential term $\exp[\beta(E-\mu)]$ in the denominator of the Fermi function is weighted by the factor $G \equiv \exp(-\beta E_n \delta)$ which strongly depends on $E_n$ and $T$. For typical magnetic fields $B$ (ranging from 0.5 Tesla to 10 Tesla) and temperatures $T$ (ranging from 1.2K to over 200K), the factor $G$ is, for practical purposes, equal to 1. However, with the temperature $T \sim 1$ mK and the magnetic field $B$ of the order of 2 Tesla, 6 Tesla or 10 Tesla, $G$ = 0.45, 0.13 and 0.03 respectively. Therefore, if $E_n$ is increased by the application of a high magnetic field ($B > 20$ Tesla) and the temperature is sufficiently lowered ($T \lesssim 5mK$) then new, interesting electronic effects may be observed provided that the condition $\mu^*(T,B;\delta) \geq \mu(T,B;0)$ is valid.

As a simple example, we will consider the electronic specific heat $C_v$ of a degenerate system of electrons held at low temperatures under high magnetic fields. In the absence of external fields, the specific heat of the electron gas is given by the well-known formula:
$$C_v = (Nm/\hbar^2)(\pi V_s/3N)^{2/3} T \qquad (3.10)$$
which shows a linear temperature dependence, vanishing only at $T = 0$. In the above expression, $m$ is the electron mass, $N$ is the number of electrons and $V_S$ is the volume of the system. In deriving eqn (3.10), the Fermi function $f(E_n) = [\exp \beta(E_n - \mu) + 1]^{-1}$ is assumed. If, however $f(E_n)$ is significantly modified as described above, it should be expected that at low temperatures and high magnetic fields, thermodynamic properties such as the specific heat will be profoundly influenced.

Consider a degenerate electron gas in a high magnetic field ($B \sim 20$ Tesla) and at low temperature ($T \sim 1$mK) such that $G \approx 10^{-3}$ for $E_n$, $n \geq 0$. At $E_n = \mu$, the probability of occupation increases from ½ to 0.9995... . For values of $E_n$ satisfying the expression $-kT + \mu \leq E_n \leq kT + \mu$, the modified Fermi function $f^*(E_n) \approx 1$, hence the temperature dependence of the occupation factor is effectively suppressed by the high magnetic field. Under this condition, the total energy of the electron gas is

$$E_T \approx D(B)(1-\delta)[\sum_{n=0}^{J}(n+1/2)\mu_B B + \hbar^2(J+1)/2m \int_0^{K_F} k_z^2 dk_z] \qquad (3.11)$$

which clearly is independent of $T$. In eqn (3.11) the sum over $n$ is truncated at $n = J$ where $E_J \gg kT + \mu$, $f(E_{J+1}) \approx 0$ but $f^*(E_{J+1}) \approx 1$ as shown (schematically) in Fig. 1. This implies that for a degenerate electron gas in the presence of a magnetic field, $C_v \equiv \partial E_T/\partial T \approx 0$ at a temperature $T \neq 0$ if $k_B T/E_0 \ll \delta$. Thus, as the magnetic field is increased in magnitude, an interval of temperature [0, $T'_R$] is observed within which $C_v \sim 0$; here $T'_R$ is of the order of a few mK.

An (arbitrary) criterion can be established for determining the range of temperature within which $C_v$ is vanishingly small by considering the Fermi function (and the modified Fermi function) at chosen pairs of temperature and magnetic field. Let $E_{n=j}$ be the energy of a filled Landau level. Let it be assumed also that at $B = 0$ and $T = T^*$, $f(E_{n=j}) = 0.25$ and that at $B \neq 0$, $T = T^*$, $f^*(E_{n=j}) = 0.75$. Equating $E_{n=j}$ from $f^*(E_{n=j})$ and $f(E_{n=j})$, we observe that $f(E_{n<j}) \sim 1$ (hence $C_v \sim 0$) when $0 < T' < (kT^* \ln 4/3 + E\delta)/k \ln 4$. Thus the range of temperatures (which includes $T^*$) for which the external field modifies the specific heat $C_v$ of the electron gas is directly proportional to the magnetic field $B$, where $k_B T^* \ll \hbar\omega\delta$. For example, if $T^* \sim 1$ mK and a field $B = 20$ T is applied, then the specific heat $C_v \sim 0$ in the range $0 < T_R' < 6.04$ mK. The near vanishing of $C_v$ may also occur at higher temperatures with lower magnetic fields if the effective cyclotron mass $m^*$ is much smaller than the free electron mass $m$.

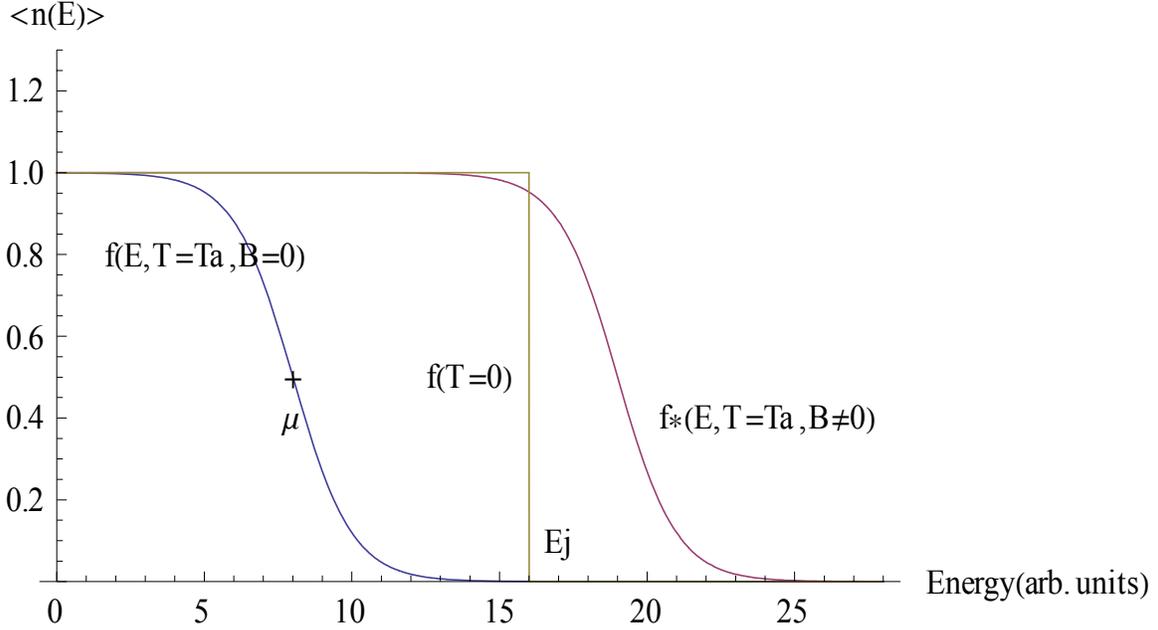

Fig.1. Under a high magnetic field and at low finite temperature $T^*$, the degenerate electron gas approximately simulates the behavior in the ground state ($T = 0$), of a system of electrons which has a chemical potential $\mu \equiv E_J > \mu^*$.

Analytically, the specific heat $C_{vm}$ of the electron gas in a magnetic field (in the absence of the perturbative correction) may be determined by differentiating with respect to $T$, a modification of the expression (3.6). Assuming that $E_T$ is weakly dependent[2] on $k_z$ within a finite range of wave-vectors $(\pm q_z \pi / 2L)$ in the z-direction and also assuming that under the magnetic fields considered the Landau levels are closely spaced, then eqn (3.6) may be re-written as

$$E_T \approx \bar{D}(B) \int_0^\infty \frac{E dE}{e^{\beta(E-\mu)} + 1} \quad ; \quad \bar{D}(B) \gtrsim D(B) \tag{3.12}$$

where the summation over $n$ has been replaced by an integration over $E$. Hence,

$$C_{vm} \equiv \frac{\partial E}{\partial T} \approx \frac{\bar{D}(B)}{2\beta} \int_{-\infty}^\infty \frac{x^2 e^x dx}{(e^x + 1)^2} \quad ; \quad x = \beta(E - \mu) \tag{3.13}$$

---

[2] The assumption that $E_T$ is independent of $k_z$ is exactly correct only for the electron gas in 2D.

Evaluating the above 'Sommerfeld integral' in eqn (3.13), it is easy to see that the specific heat,

$$C_{vm} \approx \frac{\bar{D}(B)}{6} \pi^2 k_B^2 T \qquad (3.14)$$

Thus, when the perturbation is not considered, the behaviour of the specific heat $C_{vm}$ is similar to that of the zero-field case, varying linearly with temperature and vanishing only at $T = 0$. Also, it depends linearly on the magnetic field via the density of states $\bar{D}(B)$; hence it differs only in slope from that of the free-field situation eqn (3.10).

If however, the perturbation is taken into considerations, the total electron energy $E_T$ is modified by the replacement $E \to E' = E - E\delta$. It is observed that the functional form of $E_T$ is not significantly altered, but in calculating the specific heat $C_{vp}$, the variable $x$ in the 'Sommerfeld integral' is replaced with $x' = x - \beta E\delta = ax - b$, where $a = (1-\delta)$ and $b = \beta\mu\delta$. Putting $x = x'$ in eqn (3.13), we find that the specific heat

$$C_{vp} \approx \frac{\bar{D}(B)}{2\beta(1-\delta)} \sum_{p=0}^{2} \int_{-\infty}^{\infty} e^{-b} \frac{d_p x^{2-p} e^{xa} dx}{(e^{-b} e^{xa} + 1)^2} \qquad (3.15)$$

where $d_0 = a^2, d_1 = -2ab, d_2 = b^2$ and $0 < a < 1$. It may be noticed from (3.15) that as $\delta \to 0$ (or $T \to \infty$), $b \to 0$ and only the first ($p = 0$) integral remains; also $b \to 0$ implies that $e^{-b} \to 1$, hence $C_{vp} \to C_{vm}$, i.e the effect of the perturbation disappears at high temperatures. As $T \to 0$, $b = \beta\mu\delta$ increases, and the other terms in the series contribute. The integral in eqn (3.13) with $x \to x' = x - \beta E\delta$ may be evaluated in order to show how, in general, $C_{vp}$ varies with the temperature and the magnetic field.

We shall here describe however, the equivalent result of evaluating (3.15) which shows that the curve of $C_{vp}$ has a local minimum at $T = 0$ and an intersection with the curve of $C_{vm}$ at a finite temperature $T_x$ as schematically shown in Fig. 2. These observations follow by noting that at low temperatures and high magnetic fields $e^{-b} e^{xa} \ll 1$ in the denominator of eqn (3.15), thus allowing the re-expression of the integrals in terms of the gamma functions $\Gamma(n)$ where n = 1, 2, 3. We find that,

$$C_{vp} \approx \frac{\bar{D}(B)}{(1-\delta)} (2K_B T + 2\mu\delta + \frac{\mu^2 \delta^2}{K_B T}) e^{\frac{-\mu\delta}{KT}} \equiv A(B,T)\varphi(b) e^{-b} \qquad (3.16)$$

from which may be determined, the extrema values of $C_{vp}$ as a function of temperature; it may be observed that (3.16) has no local maximum. At intermediate temperatures when $e^{-b} e^{xa} \sim 1$, the denominator of the integrand in (3.15) may be expressed approximately as $4e^{-2b} e^{2xa}$. Hence,

$$C_{vp} \approx \frac{\bar{D}(B)}{(1-\delta)} [0.5 k_B T + \frac{a\mu\delta}{2} + \frac{(\mu\delta)^2}{4 k_B T}] e^{\frac{\mu\delta}{KT}} \equiv 0.25 A(B,T)\varphi(b) e^{b} \qquad (3.17)$$

which decays exponentially as the temperature increases, tending towards the straight line given by the expression $C_{vp} \approx 0.5[\bar{D}(B)(1-\delta)kT + a\mu\delta]$. At higher temperatures $b \to 0$, $e^{-b} \to 1$ and $C_{vp} \approx C_{vm}$ (a straight line) as previously remarked, but here again given as a consequence of eqn (3.17).

The specific heat $C_{vp}$ under the conditions specified above, is described at low temperatures by an exponential curve which transforms at higher temperatures into the straight line described by the expression $C_{vm}$; the transition may occur sharply at a critical temperature $T_c$ or gradually evolve within a finite temperature interval. If, in comparison with the other terms, we neglect the constant ($a^2$) term in the function $\varphi(b)$ described above in eqn (3.17), then the approximate expression obtained describes a function $\tilde{C}_{vp}$ which has a local maximum at $T_m \approx \mu\delta/1.4 k_B$, a zero at $T = 0$ and a local minimum also at $T = 0$. For $\mu = 3\,meV$, the peak occurs at about $T_m = 30\,mK$.

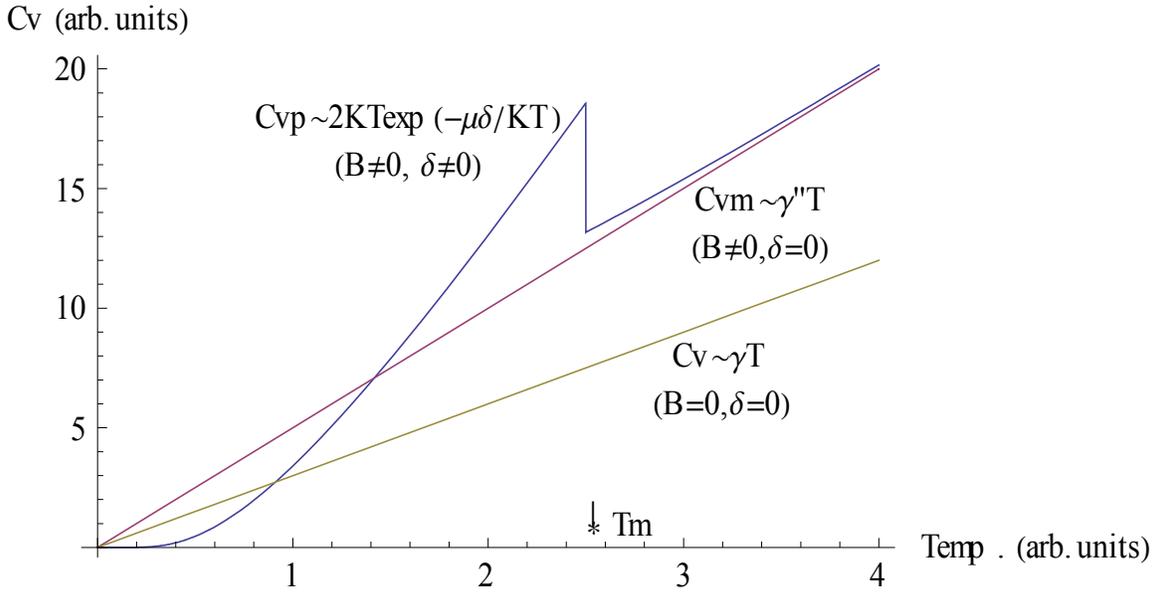

Fig.2. At low temperatures the specific heat $C_{vp}$ of the electron gas is vanishingly small, but approaches the linear law at higher temperatures. The specific heat has a discontinuity in slope at $T_m$, which indicates that a probable phase transition occurs at this temperature.

Though the above analysis gives interesting useful results, it must be mentioned that the calculations are partly approximate, because of the assumption (among others[3]) of closely spaced Landau levels which strictly is not valid at high magnetic fields. Accurate, quantitative values of $C_{vp}$ may be obtained by differentiating with respect to $T$ the eqn (3.8) and evaluating the resulting expression using numerical methods of integration, but the results obtained will not in essence, deviate from that given above.

---

[3] It is also assumed that the chemical potential µ* vary slowly with the temperature and the magnetic field.

## C. Magneto-Optical Effects in Semiconductors: Magneto-absorption

In the preceding sections, we have considered indirect ways of detecting the presence (or otherwise) in eqn (2.2) of the extra kinetic energy term $(\pi^2 \delta(p)/2m)$ which corrects the Landau levels $E_n$ and the zero-field Fermi energy $E_F$. There is however, a direct method of observing the Landau levels which may also prove useful for determining the small energy corrections $E_n \delta(p_n)$.

When placed in a magnetic field in the z-direction, a simple band semiconductor forms in the valence and conduction bands quantized (Landau) levels which are concentric parabolas described by eqn (1.3). With $k_z = 0$ and neglecting spin, the energy for the electrons and holes, when measured from the bottom of the zero-field conduction band, are $E_n = (n+1/2)\hbar\omega_{cc}$ and $E_{n'} = -\Delta E(0) - (n'+1/2)\hbar\omega_{cv}$, where $\Delta E(0)$ is the band gap in the absence of the field, and $\omega_{cv}$ & $\omega_{cc}$ are the electron cyclotron frequencies for the valence and conduction bands respectively. To make optically-induced transitions from the valence to the conduction band the photon energy $\hbar\omega$ required, when the correction to the Landau levels is considered, is given by

$$\hbar\omega = \Delta E(0) + (n+1/2)\hbar(\omega_{cc} + \omega_{cv}) - (n+1/2)\hbar(\omega_{cc} + \omega_{cv})\delta \; ; \; n' = n \qquad (3.18)$$

the selection rules are $\Delta n = 0, \Delta k_z = 0$. In (3.18) $\omega_{cc} = eB/m_c c$, $\omega_{cv} = eB/m_v c$ and ($m_v$, $m_c$) are the cyclotron effective masses in the valence and conduction bands respectively.

The absorption coefficient for the allowed transitions between the Landau levels in the valence and conduction bands is

$$\tilde{\alpha}(B) = \frac{A_\gamma \omega'_c}{2\omega} \sum \frac{1}{(\omega - \omega_n)^{1/2}} \qquad (3.19)$$

where the quantity $A_\gamma$ is proportional to the momentum matrix $|\mathbf{P}_{cv}|$. Thus the measured absorption coefficient will exhibit singularities or peaks [16] for transitions between the quantized levels. This has been observed in transmission measurements on single crystals of Ge, GaAs, InSb and other semiconductors at temperatures as low as 4K and magnetic fields as high as 10Tesla.

The quasi-continuous distribution of states in the z-direction together with the possible indirect transitions cause the appearance of marked peaks instead of lines on the high-energy side of the fundamental absorption peak. These oscillatory peaks are equally spaced along the energy scale, their spacing being directly proportional to the magnetic induction and inversely proportional to the reduced cyclotron mass $m'$. The appearance of the peaks constitutes an immediate proof of the existence of the Landau levels and the measured shift from that expected based on the expression $\hbar\omega = \Delta E(0) + (n+1/2)\hbar\omega'_c$, of the photon frequencies at which these peaks occur, will indicate the presence (or otherwise) of the correction described in (3.18).

## 4. Experimental Considerations and Discussion of Results

At temperatures and magnetic fields satisfying the condition $k_B T \gg \hbar \omega_c$, the magnetization (or diamagnetic susceptibility) of a system of electrons is steady. However, at low temperatures ($T \leq 1.2K$) and high magnetic fields ($B > 15$ T) satisfying the condition $\hbar \omega_c \gg k_B T$, the magnetization (or susceptibility) of several metals oscillate as a function of $1/B$. This is the de Haas-van Alphen effect (dHvA), which appears as a consequence of the quantized Landau levels. In a magnetic field, the total electron energy (and several other properties) of the non-interacting electron gas oscillate at low temperatures at a frequency given by the Onsager relation $f \propto A(E_F)$ where $A$ is the area of the extremal cross-section of the Fermi surface, $E_F$ is the zero-field Fermi energy and $f$ & $A$ are related by a universal constant. As demonstrated in section 3A, the frequency of the dHvA oscillations is however changed by the presence of an intrinsic perturbation which modifies both the Landau levels and the zero-field Fermi energy level. It is expected that the modification will become observable at temperatures and magnetic fields satisfying the condition $\hbar \omega_c \delta \gg k_B T$; hence the frequency correction given by eqn (3.4) will remain fixed as the temperature and magnetic fields are changed to satisfy this condition.

A possible method of testing the above conclusions is by carrying out, on non-superconducting solids with well-known Fermi surfaces, the de Haas-van Alphen measurement first at a typical field (10 Tesla) and temperature ($T \sim 1.2$ K) and then at very low temperatures ($T \sim 5$ mK) and higher magnetic fields ($B \sim 25$ Tesla). Some metals (in groups 1A, 2A and 1B of the periodic table) with large zero-field Fermi energy such as Cu(7.04eV), Ag(5.48eV), Au(5.51eV) appear suitable to the observation of the effects described above. Also, materials with ellipsoidal Fermi surfaces with strong mass anisotropy in the direction of the magnetic field should prove interesting. If there are solids in which it is found that for certain field orientation or some other variable $(\partial S / \partial E)|_\eta$ much smaller (or much greater) than 1, then it will be possible to establish, in a very striking way, the correction to the frequency (or period) of oscillation of the magnetization (or susceptibility).

The behavior of the electronic specific heat $C_v$ of a degenerate Fermi gas in a high magnetic field is also modified as a result of the correction of the Landau levels. It is found that the specific heat at low temperatures is anomalous; from absolute zero, it increases with temperature exponentially and at high temperatures it coincides with a straight line. To observe this effect, the temperature and magnetic fields should be such that $E_n \delta \sim \mu \delta \sim 8 k_B T$. Also, the chemical potential $\mu$ should be of the order of a few milli electron-volts (~3meV - 30meV). This is obtainable with a dilute electron gas of density $N/V \sim 10^{21}/m^3$ given that $E_F \propto (N/V)^{3/2}$. Hence the 'free electron'-like alkali metals which have electron densities of the order $(N/V \sim 10^{28}/m^3)$ may not be suitable for the detection of the anomaly in the electronic specific heat of degenerate electron systems as described above. It may be required that low-density electron systems be fabricated in order to observe the effect.

The behaviour of the specific heat as shown in Fig. 2 indicates a possible phase change and/or transition to a correlated state in the system of electrons. Anomalous variation in the specific heat of a many-particle system has been associated with the onset of superconductivity in a homogeneous electron gas or the transition to superfluidity in a system of $^4$He atoms. The behaviour of $C_{vp}$ in eqns (3.16) and (3.17) is similar in some respects to that of these systems. The exponential growth from $T = 0$ of the electronic specific heat $C_v$ is also typical of a gapped state. The contraction of the Fermi surface described by eqn (2.7) is suggestive of an attractive interaction.

It may be possible to correctly determine by the well-known methods of statistical thermodynamics and measurements, the order of the phase transition and the transition temperature if we consider exact expressions for the free energies of the electron gas interacting with a magnetic field. Also significant for the above considerations is the determination of the chemical potential as a function of the temperature and magnetic field such that the condition $\mu^*(T, B; \delta) \geq \mu(T, B; 0)$ is satisfied over the appropriate range on temperatures and at a given intensity of the magnetic field. At $T = 0$ the equation, $N = (Vs\hbar\omega_c / 2\pi^2)(2m^*/\hbar^2)^{3/2} \sum_{n=0}^{n_{max}} [E_F - \hbar\omega_c(n+1/2)]^{1/2}$ relates the Fermi level $E_F$ to the number of electrons $N$ and the magnetic field $B$. A numerical solution to the equation shows that $E_F$ oscillates about the zero-field Fermi energy level $E_{F0}$ [15]. If it is assumed that the relation is only slightly changed at very low temperatures, then it is readily observed that the constraint placed on the chemical potential can be satisfied.

From the eqn (3.18) semiconductors[4] where $\omega_{cc}$ and $\omega_{cv}$ are high appear best suited for measuring the frequency shift. A choice of semiconductors with appropriate band gaps $\Delta E(0)$ may be required in order to locate the absorption frequencies in a region of the electromagnetic spectrum where highly precise measurements of frequency shifts can be made using interferometers. For ZnO, the fundamental absorption frequency lie in the ultraviolet region ($10^{15} - 10^{16}$), while for InSb it lies in the infra-red part of the spectrum ($10^{12} - 10^{14}$). Also in InSb, the band effective mass $m^{**} = 0.015m$ is approximately equal to the cyclotron effective mass $m^*$, hence the zero-point energy and the temperature at which quantum effects are observable is increased by a factor of 67. For a sample of thickness L ≈ 65 μm oriented in the direction of the magnetic field, it follows from the expression $k_B T \ll (2\hbar^2\pi^2/m^* L^2)$ that if $T \ll 70\,\text{mK}$, then it may be possible to obtain sharp peaks or discrete lines in the absorption spectrum corresponding to the Landau levels because below this temperature, thermal transitions along $k_z$ are negligible. Also, given the condition $k_B T \ll \hbar\omega\delta$ it follows that below 8 mK in a field of 20 Tesla, the corrections to the Landau levels (which are of the order $10^{-3}\hbar\omega_c$) should manifest.

We have assumed above that the independent electron approximation remains applicable. Other properties (in a high magnetic field) of a dilute, degenerate electron gas which are dependent on the Fermi function e.g. the diamagnetic/paramagnetic susceptibility, transport properties and thermodynamic properties, will be modified as a result of the change in the Landau levels due to the perturbation ($\delta \neq 0$), which manifests at high magnetic fields and ultra-low temperatures. The quantity $\delta(p; \mathbf{A})$ changes with the temperature and magnetic field and is strongly influenced by $\bar{\alpha}$ which, contrary to our previous assumption, may be different from 1 depending on the ratio $(e\mathbf{A}/pc)$; for high fields and low temperatures $\bar{\alpha} = [1 - e\mathbf{A}.\mathbf{p}/p^2c + e^2A^2/p^2c^2]^{-1} \neq 1$.

---
[4] For GaP, ΔE(0) is 2.40eV and m*= 0.5m;  InSb ΔE(0) is 0.26eV, m*= 0.013m;  ZnO, ΔE(0) is 3.2eV and m*=0.27m.

## 5. Conclusions

At very low temperatures ($T \sim 10^{-3}$ K) and high magnetic fields ($B \sim 300$ kG) corrections to the de Haas-van Alphen (dHvA) frequency in non-superconducting metals may be observed. The change in the dHvA frequency is determined by the zero-field Fermi energy, the measurable anomalous part of the electron magnetic moment $\delta(p)$, and the change (with energy) of the extremal cross-sectional area of the Fermi surface perpendicular to the magnetic field. The predicted frequency modification is a consequence of the perturbative correction (in a high magnetic field) to the Landau levels, and of the zero-field Fermi surface. Also, with suitable semi-conductors direct observation of the corrected Landau levels may be made using absorption spectroscopy at very low temperatures. High precision measurement of the frequency associated with direct interband transitions should reveal the change in the energy levels in simple-band semiconductors. Finally as a result of the intrinsic perturbation, the possibility of observing in a dilute gas of electron a phase transformation is demonstrated. The calculated specific heat exhibits anomalies which may indicate in a degenerate Fermi gas, a change of state at low temperatures and under high magnetic fields.

**Acknowledgment**

The author acknowledges with gratitude useful discussion with many colleagues.

**Appendix I: Many-body Corrections and the Lifshitz-Kosevich Theory** [1, 14].

The effects of many-body interactions do not drastically alter the structure of the Lifshitz-Kosevich formula, although they may change the parameters contained therein; the electron-electron interaction changes the Fermi surface and the g-factor, but only slightly changes the effective cyclotron mass, while the electron-phonon interaction changes only the mass. The influence on the dHvA frequency of the corrected Landau levels may nevertheless be isolated from that of the many-body effects by varying the field and temperature to satisfy the condition $\hbar\omega\delta \gtrsim k_B T$; the frequency is corrected as expressed by (3.4) only, if the field and temperature both satisfy this constraint.

If the independent particle approximation remains valid at the low temperatures & high magnetic fields herein considered, then the Lifshitz-Kosevich formula may be assumed correct. We have here taken as valid the Lifshitz-Kosevich theory (down to the lowest temperatures and up to the highest magnetic fields available) and calculated the frequency change as a consequence of the interaction which corrects the Landau levels and reduces the zero-field Fermi surface. In the alternative, the relation (2.6) could have been used to correct the thermodynamic potential $\Omega = -k_B T \sum \ln[1 + \exp\beta(\mu - E_n)]$ such that $\Omega \to \Omega^* = -k_B T \sum \ln[1 + \exp(\beta E_n \delta)\exp\beta(\mu - E_n)]$, where there now appears a factor $G^{-1} = \exp(\beta E_n \delta)$ multiplying the exponential term $\exp(\mu - E_n)/k_B T$. Calculations using $\Omega^*$ will indicate (where the pre-factor $G^{-1}$ becomes very significant), the existence (or otherwise) of the quantum oscillations at low temperatures/high magnetic fields and furthermore facilitate the determination (should the effects remain observable), the modified dHvA frequency and amplitude. Other thermodynamic properties which usually are derived as consequences of $\Omega$ will be similarly affected. Thus, experiments at temperatures $T \lesssim 10$ mK and magnetic fields $B \gtrsim 30$ Tesla will reveal with sufficient clarity any deviation from the Onsager relation the frequency of the dHvA oscillations, and plainly validate the Lifshitz-Kosevich theory upon which it is based.

**Appendix II: Derivation of the Corrected Landau Equation**

The well-known Pauli equation is the non-relativistic limit (in the lowest-order approximation) of the Dirac equation coupled to an external magnetic field. Likewise the modified Dirac equation when coupled to an external magnetic field yields, in the lowest order of approximation, the Pauli equation with some extra terms. A corrected Landau equation emerges as a special case (with $\hat{\mathbf{S}} = 0$) of the modified Pauli equation.

The extra term in the Pauli equation is the magnetic analog of the Darwin term which, as a consequence of the fluctuation of the coordinates ("zitterbewegung"), produces an observable shift in the s-levels of hydrogen-like atoms; likewise the corrections to the Pauli Hamiltonian are the consequences of the fluctuation of the rest energy which manifest as observable shifts in the energy of the Landau levels.

To derive the corrected Landau equation, we consider first the Dirac Hamiltonian $H_D = c\hat{\boldsymbol{\alpha}} \cdot \hat{\mathbf{p}} + \hat{\beta}mc^2$, to which has been added a time-dependent term which describes the fluctuation of the rest energy [7]:

$$H = c\hat{\boldsymbol{\alpha}} \cdot \hat{\mathbf{p}} + \hat{\beta}mc^2 - \hat{\beta}mc^2 \exp(2i\hat{\boldsymbol{\alpha}} \cdot \hat{\mathbf{p}} ct/\hbar) \tag{A.1}$$

The right hand side of the above equation can be expressed in a time-independent form by considering the expression;

$$\lim_{t \to \infty} \frac{1 - e^{ixt}}{x} = -\lim_{\varepsilon \to 0^+} i \int_0^\infty e^{i(x+i\varepsilon)t'} dt' \tag{A.2}$$

where $\varepsilon$ is a small positive quantity which tends to zero along the horizontal ($x$) axis and has the dimensions of inverse time $t^{-1}$, if $t$ is interpreted as the time variable. Expanding the right-hand-side,

$$\lim_{t \to \infty} \frac{1 - e^{ixt}}{x} = \lim_{\varepsilon \to 0^+} \left[ \frac{x}{x^2 + \varepsilon^2} - \frac{i\varepsilon}{x^2 + \varepsilon^2} - \frac{e^{ixt}e^{-\varepsilon t}}{x + i\varepsilon} \Big|^\infty \right] \tag{A.3}$$

Simplifying the expression by assuming that the rate at which $t \to \infty$ is faster than that at which $\varepsilon \to 0$,

$$\lim_{t \to \infty} \frac{1-e^{ixt}}{x} = \lim_{\varepsilon \to 0} \left[ \frac{x}{x^2 + \varepsilon^2} - \frac{i\varepsilon}{x^2 + \varepsilon^2} \right] \tag{A.4}$$

The expression obtained can be further simplified as follows:

$$\lim_{t \to \infty}(1 - e^{ixt}) = \lim_{x \to \infty} \left[ \frac{x^2}{x^2 + \varepsilon^2} - \frac{i\varepsilon x}{x^2 + \varepsilon^2} \right] \tag{A.5}$$

If the variables $x$ and $t$, and the parameter $\varepsilon$ are such that $x \to 0^+$ as $t \to \infty$, and $\varepsilon$ is of the same order of magnitude as $x$, i.e. $\varepsilon \sim x$, then

$$\lim_{x \to 0^+} \left[ \lim_{t \to \infty} e^{ixt} \right] \approx \lim_{x \to 0^+} \left[ \lim_{\varepsilon \to 0^+} \frac{i\varepsilon}{2x} + \frac{\varepsilon^2}{2x^2} \right] \tag{A.6}$$

These conditions can be satisfied for typical values of $\hat{\boldsymbol{\alpha}} \cdot \hat{\mathbf{p}}c$. Thus if we put $x = 2\hat{\boldsymbol{\alpha}} \cdot \hat{\mathbf{p}}c / \hbar$, we obtain

$$e^{(2i\hat{\boldsymbol{\alpha}} \cdot \hat{\mathbf{p}}ct)/\hbar} \approx \lim_{\varepsilon \to 0^+} \left[ \frac{i\varepsilon\hbar/c}{4(\hat{\boldsymbol{\alpha}} \cdot \hat{\mathbf{p}})} + \frac{1}{2} \frac{\hbar^2 \varepsilon^2}{(2\hat{\boldsymbol{\alpha}} \cdot \hat{\mathbf{p}}c)^2} \right] \tag{A.7}$$

as $t \to \infty$ and $x \to 0^+$.

Inserting (A.7) in (A.1) and noting that the term reciprocal in $(\hat{\boldsymbol{\alpha}} \cdot \hat{\mathbf{p}}c)^2$ is small compared with $(\hat{\boldsymbol{\alpha}} \cdot \hat{\mathbf{p}}c)$ and $(\hat{\beta}mc^2)$, and may therefore be neglected, the Hamiltonian takes the time-independent form:

$$H = c\hat{\boldsymbol{\alpha}} \cdot \hat{\mathbf{p}} + \hat{\beta}mc^2 - \frac{\hat{\beta}mci(\hbar\varepsilon)}{4(\hat{\boldsymbol{\alpha}} \cdot \hat{\mathbf{p}})} \tag{A.8}$$

where $\hbar\varepsilon$ is interpreted as the energy of a massless excitation which propagates in association with the electron. Using the identity $(\hat{\boldsymbol{\alpha}} \cdot \hat{\mathbf{p}})(\hat{\boldsymbol{\alpha}} \cdot \hat{\mathbf{p}}) = \hat{\mathbf{p}} \cdot \hat{\mathbf{p}}$, we can reduce the above to

$$H = c\hat{\boldsymbol{\alpha}} \cdot \hat{\mathbf{p}} + \hat{\beta}mc^2 - \frac{\hat{\beta}mci(\hbar\varepsilon)(\hat{\boldsymbol{\alpha}} \cdot \hat{\mathbf{p}})}{4|\hat{\mathbf{p}}|} \tag{A.9}$$

The time-independent eigenvalue equation corresponding to the above Hamiltonian is

$$E_n \psi_n = \left[ c\hat{\boldsymbol{\alpha}} \cdot \hat{\mathbf{p}} + \hat{\beta}mc^2 - \frac{(\hat{\beta}\hat{\boldsymbol{\alpha}} \cdot \hat{\mathbf{p}})mci(\hbar\varepsilon)}{4|\hat{\mathbf{p}}|} \right] \psi_n \tag{A.10}$$

We require that for the eigenfunctions of $H$ the superposition principle be satisfied, hence we linearize the Hamiltonian by noting first that $|\hat{\mathbf{p}}| = \hat{\mathbf{p}} \cdot \hat{\mathbf{p}}$ commutes separately with $H_D = c\hat{\boldsymbol{\alpha}} \cdot \hat{\mathbf{p}} + \hat{\beta}mc^2$ and $\hat{\beta}\hat{\boldsymbol{\alpha}} \cdot \hat{\mathbf{p}}$; thus we can rewrite (A10) by using the fact that $|\hat{\mathbf{p}}|\psi_n = |\mathbf{p}_n|\psi_n$, where $\mathbf{p}_n$ is the eigenvalue in the state $\psi_n$ of the free-particle momentum operator $\hat{\mathbf{p}}$. Hence we have,

$$E_n \psi_n = \left[ c\hat{\boldsymbol{\alpha}} \cdot \hat{\mathbf{p}} + \hat{\beta}mc^2 - \frac{(\hat{\beta}\hat{\boldsymbol{\alpha}} \cdot \hat{\mathbf{p}})mci(\hbar\varepsilon)}{4|\mathbf{p}_n|} \right] \psi_n \tag{A.11}$$

In an electromagnetic field the equation (A.11) becomes, by minimal coupling,

$$E_n\psi_n = \left[c\hat{\boldsymbol{\alpha}}\cdot\hat{\boldsymbol{\pi}} + \hat{\beta}mc^2 - \frac{(\hat{\beta}\hat{\boldsymbol{\alpha}}\cdot\hat{\boldsymbol{\pi}})mci(\hbar\varepsilon)}{4|\mathbf{p}_n - e\mathbf{A}/c|}\right]\psi_n \tag{A.12}$$

Expanding the denominator of the third term in the square brackets and factoring out the momentum $p_n^2$ (i.e the magnitude of $\mathbf{p}$), we have $p^2[1 - 2e\mathbf{A}\cdot\mathbf{p}/p^2c + (eA/p)^2]$. The equation (A.12) above can be simplified by putting $\bar{\alpha} = [1 - 2e\mathbf{A}\cdot\mathbf{p}/p^2c + (eA/pc)^2]^{-1}$ such that,

$$E_n\psi_n = \left[c\hat{\boldsymbol{\alpha}}\cdot\hat{\boldsymbol{\pi}} + \hat{\beta}mc^2 - \frac{(\hat{\beta}\hat{\boldsymbol{\alpha}}\cdot\hat{\boldsymbol{\pi}})\bar{\alpha}mci(\hbar\varepsilon)}{4p_n^2}\right]\psi_n \tag{A.13}$$

We shall find the non-relativistic limit of the above by substituting into it the Dirac bi-spinor, we obtain the two equations

$$E_n\psi_{nA} = c\hat{\boldsymbol{\sigma}}\cdot\hat{\boldsymbol{\pi}}\psi_{nB} + mc^2\psi_{nA} - \frac{i\hbar m\bar{\alpha}c\varepsilon(\hat{\boldsymbol{\sigma}}\cdot\hat{\boldsymbol{\pi}})}{4p_n^2}\psi_{nB} \tag{A.14a}$$

$$E_n\psi_{nB} = c\hat{\boldsymbol{\sigma}}\cdot\hat{\boldsymbol{\pi}}\psi_{nA} - mc^2\psi_{nB} + \frac{i\hbar m\bar{\alpha}c\varepsilon(\hat{\boldsymbol{\sigma}}\cdot\hat{\boldsymbol{\pi}})}{4p_n^2}\psi_{nA} \tag{A.14b}$$

which when (A.14b) is re-written reduces to

$$\psi_{nB} = \frac{c(1 + i\hbar m\varepsilon\bar{\alpha}/4p_n^2)}{E + mc^2}(\hat{\boldsymbol{\sigma}}\cdot\hat{\boldsymbol{\pi}})\psi_{nA} \tag{A.15}$$

Putting $E_n - mc^2 = E_{nA}$ we write (A.14a) as

$$E_{nA}\psi_{nA} = c(1 - i\hbar m\bar{\alpha}c/4p_n^2)(\hat{\boldsymbol{\sigma}}\cdot\hat{\boldsymbol{\pi}})\psi_{nB} \tag{A.16}$$

Substituting $\psi_{nB}$ in (A.15) and (A.16) we obtain

$$E_{nA}\psi_{nA} = \frac{c^2(1 + \hbar^2m^2c^2\bar{\alpha}^2/16p_A^4)(\hat{\boldsymbol{\sigma}}\cdot\hat{\boldsymbol{\pi}})^2}{E + mc^2}\psi_{nA} \tag{A.17}$$

We note that $(\hat{\boldsymbol{\sigma}}\cdot\hat{\boldsymbol{\pi}})(\hat{\boldsymbol{\sigma}}\cdot\hat{\boldsymbol{\pi}}) = \hat{\boldsymbol{\pi}}\cdot\hat{\boldsymbol{\pi}} + (\hat{\boldsymbol{\sigma}}\cdot\mathbf{B})(e\hbar/c)$ and $E + mc^2 \approx 2mc^2$ because $E \approx mc^2$. Hence,

$$E_{nA}\psi_{nA} = \frac{[1 + \hbar^2m^2\varepsilon^2\bar{\alpha}^2/16p_A^4][\hat{\boldsymbol{\pi}}^2 - (e\hbar/c)(\hat{\boldsymbol{\sigma}}\cdot\mathbf{B})]}{2m}\psi_{nA} \tag{A.18}$$

Expanding (A.18) we obtain

$$E_{nA}\psi_{nA} = \left[\frac{\hat{\boldsymbol{\pi}}^2}{2m} - \frac{\hbar^2m^2\hat{\boldsymbol{\pi}}^2\varepsilon^2\bar{\alpha}^2}{2m4^2p_A^4} - \frac{e\hbar}{2mc}\left(1 + \frac{\hbar^2m^2\varepsilon^2\bar{\alpha}^2}{4^2p_A^4}\right)(\hat{\boldsymbol{\sigma}}\cdot\mathbf{B})\right]\psi_{nA} \tag{A.19}$$

For $\mathbf{S} \equiv \hbar(\boldsymbol{\sigma}/2) = 0$, we have (dropping the subscript n)

$$E_A\psi_A = \left[\frac{\hat{\boldsymbol{\pi}}^2}{2m} - \frac{\hat{\boldsymbol{\pi}}^2\hbar^2m^2\varepsilon^2\bar{\alpha}^2}{2m(4^2p_A^4)}\right]\psi_A \tag{A.20}$$

The physical basis of the correction to the Landau equation is the fluctuation of the rest energy which, together with the "zitterbewegung" and the other relativistic quantum fluctuations, all have as their common origin the interference of the positive and the negative energy states.